# Agarose Derived Carbon Based Nanocomposite for Hydrogen Storage at Near-Ambient Conditions

*A Flamina[1], R M Raghavendra[2], Anandh Subramaniam[1] and Raghupathy Yuvaraj[1,*]*

[1]Materials Science and Engineering, Indian Institute of Technology, Kanpur, India.

[2]Applied Mechanics and Biomedical Engineering, Indian Institute of Technology, Madras, India.

*Author for correspondence: raghu@iitk.ac.in*

**Abstract**

Nanocomposites comprising of high surface area adsorption materials and nanosized transition metals have emerged as a promising strategy for hydrogen storage application due to their inherent ability to store atomic and molecular forms of hydrogen by invoking mechanisms like 'physisorption' and 'spillover mechanism' or 'Kubas interaction'. The potential use of these materials for both transport and stationary applications depends on reaching the ultimate storage capacity and scalability. In addition to achieving good hydrogen storage capacity, it is also vital to explore novel and efficient synthesis routes to control the microstructure. Herein, a direct and simple thermal decomposition technique is reported to synthesize carbon-based nanocomposites, where nickel nanoparticles are dispersed in a porous carbon matrix. The structure, morphology, composition and nature of bonding in the samples were investigated using transmission electron microscopy, scanning electron microscopy, energy dispersive spectroscopy, X-ray diffraction and Raman spectroscopy. Sorption-desorption isotherms were used to study the hydrogen storage capacity of the nanocomposites at a moderate $H_2$ pressure of 20 bar. Among the various nanocomposites examined, the best obtained storage capacity was 0.73 wt.% (against 0.11 wt.% for pure carbon sample) at 298 K with reversible cyclability. It is shown that the uniform dispersion of catalytic nanoparticles along with a high surface area carbon matrix helps in the enhancement of hydrogen storage capacity by a factor of 6.5 times over pure carbon.

## 1. Introduction

Since the 1973 energy crisis, the need to replace the existing fossil-fuel based energy economy has become crucial and emphasis on hydrogen as a potential energy carrier has gained considerable stimulus [1]. Despite huge benefits, the transition to a hydrogen economy has several setbacks starting from the production to the utilization stage. In particular, hydrogen storage is considered a major impediment to the integration of hydrogen with the global energy economy due to its low volumetric energy density (0.089 g/L at Standard Temperature Pressure) [2,3]. Compared to the expensive and unsafe high-pressure compressed gas cylinders (700 bar) and energy intensive cryogenic liquefaction storage (21 K), solid-state hydrogen storage is recognized as a more feasible and viable option. In solid-state storage, hydrogen can interact with the material either by physisorption (via weak Van der Waals force), chemisorption (via strong chemical bond) mechanism or quasi-molecular binding (intermediate between physisorption and chemisorption) [4]. Due to their unique structural stability and tunability, physisorbents like carbonaceous materials are intriguing choices for hydrogen storage. Although these materials exhibit good reversibility and kinetics, due to their low binding energy (<10 kJ/mol), they show poor storage capacity at ambient conditions [5]. Unfortunately, none of the current materials in the open literature are considered suitable for commercialization as they fail to meet the necessary DOE targets of achieving an ultimate gravimetric energy density of 0.065 kg $H_2$/kg system (6.5 wt%) at an operating temperature of −40 °C to 60 °C [6]. Additionally, the existing lab-scale procedures utilized to synthesize these materials must be less complicated and scalable to lower the overall cost of the storage system (ultimate target $266 per kg/$H_2$) [7]. So far, the hydrogen uptake of pure adsorbents at ambient temperature (298 K) is between 0.6 to 0.8 wt.% at 100 atm [8,9]. It has therefore become a subject of extensive research across the globe to explore new materials or to improve the performance of existing materials.

Nanostructuring is a promising technique to improve the gravimetric storage capacity at ambient operating conditions wherein the advantages of physisorption and chemisorption materials can be combined simultaneously via tailoring the microstructure at nanoscale [10-19]. Some of the approaches employed are (i) the use of nanohybrids, where the binding energy of the pure carbon material is improved through functionalization or doping with light heteroatoms or catalytic nanoparticles [11,12], (ii) the use of nanocontainers, where core-shell/hollow nanostructures are engineered to store both atomic and molecular form of hydrogen inside and on the surface [13-16], (iii) nanosizing/nanoscaling, where the particle size/pore space of the sorption material is altered to enhance the hydrogen adsorption enthalpy [17] and (iv) nanoconfinement where an active material is constricted within a porous matrix of high surface area material like carbon, which apart from providing structural support, also helps in preventing the growth, agglomeration and oxidation of the confined nanoparticles [18-21].

Among the strategies listed above, the emerging nanoconfinement concept is gaining prominence [20,21]. By optimizing the synthesis parameters, the particle size and dispersion of the metallic nanoparticles can be effectively controlled. Furthermore, the intrinsic challenge of the nanoscaling technique wherein the support-free nanosized material tends to agglomerate during the hydrogen sorption can be eliminated. By infiltrating nanoporous carbon with molten magnesium, de Jongh et al. prepared carbon supported non-oxidized magnesium crystallites of 2-5 to less than 2 nm size [22]. Zhang et al. incorporated $MgH_2$ in a carbon scaffold (matrix) and reported a 5 times faster dehydrogenation rate than ball milled $MgH_2$ with graphite and attributed the observation to smaller particle size of the former when compared to the latter [23]. Based on the type of confined nanoparticle, two mechanisms are possible i.e. (i) Spillover mechanism or (ii) Kubas interaction [24]. The addition of transition metal nanoparticles to carbon nanostructures is expected to increase the reversible hydrogen storage capacity at room temperature via spillover mechanism, in which the hydrogen molecule gets dissociated at the metal surface, followed by its migration and adsorption on the carbon surface [25, 26]. M. Zielinski et al. reported a sharp increase in hydrogen uptake at room temperature for nickel catalyst supported activated carbon (0.29 wt.% against 0.12 wt.% for activated carbon at 20 bar) [27]. Zhong et al. observed that the storage capacity of Pt doped carbon aerogel increased by a factor of four i.e. 0.43 wt.% at 298 K and 20 bar when compared to undoped carbon aerogel (0.1 wt.%) [28]. The hydrogen storage capacity increased from 0.18 wt.% to 0.55 wt.% after doping Ni in activated carbon nanofibers at 20 bar and 298 K [29]. At 298 K and 5.5 MPa, the hydrogen uptake of cobalt embedded ordered mesoporous carbon (OMC) was 0.45 wt%, while pure OMC adsorbed 0.2 wt.% [30]. Nevertheless, the many approaches followed for the synthesis of these materials are not scalable and the procedures followed are complex [10]. In addition, material cost is another major factor which decides the viability of the technology. In this regard, the replacement of the tedious multi-step synthesis protocols with facile, low cost synthesis protocols is critical.

The current work is in line with meeting the DOE targets and moving towards bulk production, both of which are yet to be realized. The objective of the work is to increase the sorption capacity at moderate conditions by combining the mechanisms of (i) physisorption of molecular hydrogen, and (ii) chemisorption of atomic hydrogen through spillover caused by transition metal. Here, we present a simple thermal decomposition synthesis route to fabricate a carbon-nickel (C-Ni) nanocomposite using relatively inexpensive precursors. Nickel acetate tetrahydrate (NiAc) is used as a precursor and the carbon derived from a biocompatible polysaccharide namely agarose is used as a matrix. An interesting microstructure of thermally decomposed NiAc shows nickel nanoparticles surrounded by carbon [31]. By taking advantage of its catalytic activity, this microstructure is adopted to assist in the spillover mechanism and improve the hydrogen storage capacity in carbon-based materials. Though NiAc can function as a sole precursor of both carbon and nickel, one other reason to use agarose is to increase the carbon content and optimize the Ni particle size [32,33]. The catalytic activity of nickel nanoparticles is often reduced due to their clustering and surface oxidation when exposed to the atmosphere or prolonged storage [34,35]. The use of surfactants to circumvent the sintering problem has proved to be expensive as well as a hindrance for scaling up the synthesis conditions eventually affecting their end application [36]. In this study, we report that the carbon layer coating the Ni nanoparticle protects them against environmental degradation. Furthermore, the well-defined, porous high surface area carbon matrix is believed to be an effective material to improve the hydrogen storage properties by providing structural stability to the confined material and preventing its agglomeration. In addition, it also acts as active adsorption sites for hydrogen molecules.

## 2. Experimental methodology

A simple four-step synthesis protocol was carried out to prepare the C-Ni nanocomposite (Fig. 1): (i) Preparation of precursor solutions, (ii) Formation of hydrogel, (iii) Freeze drying of the hydrogel to get aerogel and (iv) Thermal decomposition of aerogel to obtain C-Ni nanocomposite. Chemicals used in this study are Nickel Acetate tetrahydrate, NiAc (Ni(CH$_3$.COO)$_2$.4H$_2$O, Loba Chemie Pvt. Ltd., 98%), Agarose (C$_{24}$H$_{38}$O$_{19}$, s.d. fine-chem Ltd.) and de-ionised water (DI) without further purification.

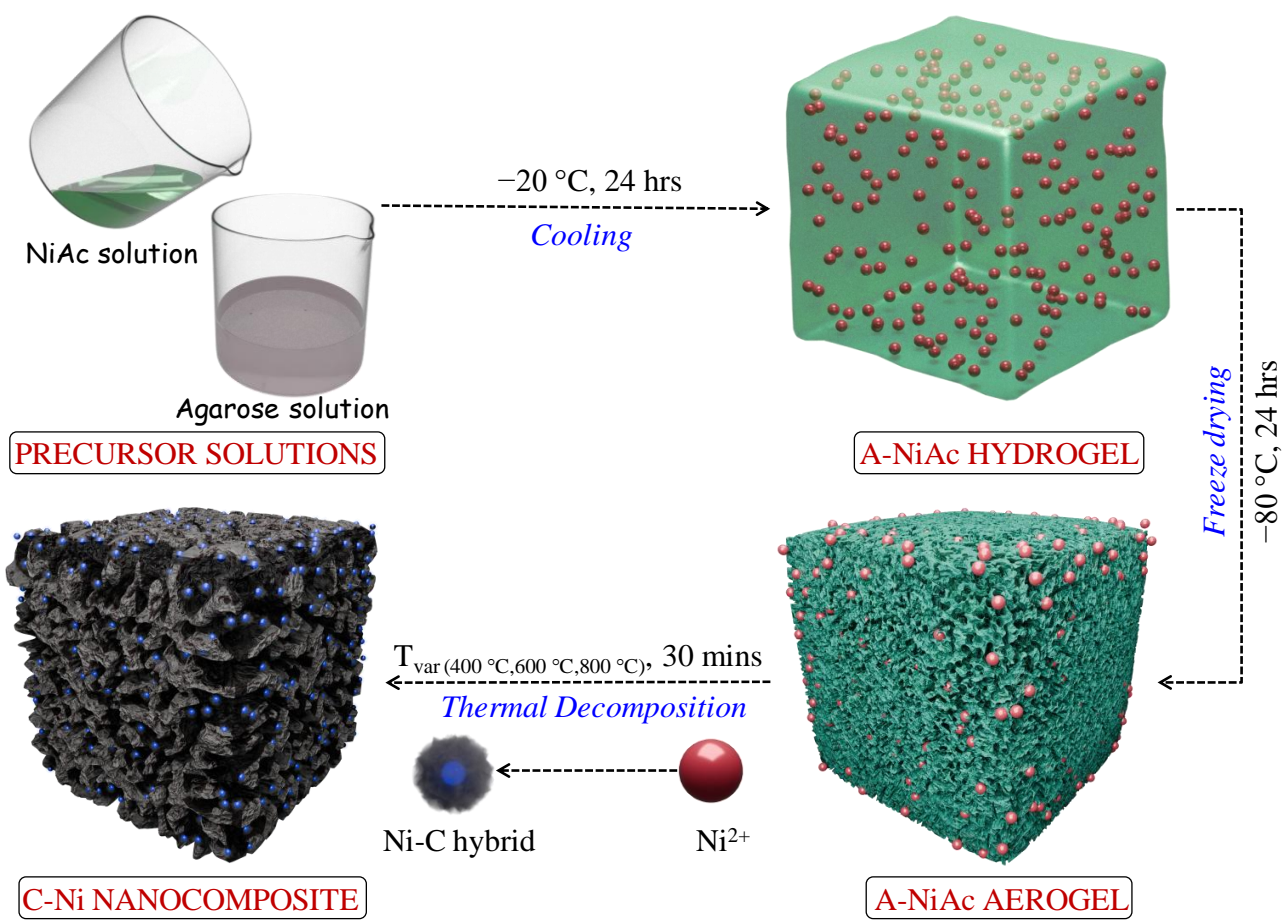

Fig. 1. Schematic illustration of the steps involved in the preparation of C-Ni nanocomposite. The thermal decomposition of NiAc precursor (Ni$^{2+}$) to Ni nanoparticle covered with a thin layer of carbon is also shown.

The detailed synthesis procedure is as follows: 0.8 g of the agarose was mixed in 20 ml DI and heated at 90 °C under continuous stirring till the solution became transparent and clear. To nanoconfine Ni nanoparticles (Ni-C), the NiAc precursor solution was prepared and added to the agarose solution in the weight ratios of 1:10. After 30 mins of stirring, the final solution was refrigerated for 24 hrs at −20 °C to form Agarose-NiAc (A-NiAc) hydrogel. The hydrogel thus formed was freeze-dried in a lyophilizer for 24 hrs to obtain a dehydrated sponge-like structure (labelled A-NiAc aerogel) which was subsequently heat treated in argon atmosphere to produce C-Ni nanocomposites. To get an insight into the nucleation of nickel nanoparticles in the nanocomposite, the samples were thermally decomposed at various temperatures of 400 °C, 600 °C and 800 °C for 30 mins at a ramping rate of 5 °C/min (based on thermogravimetric analysis (TGA) results). The heat-treated A-NiAc aerogel was labelled as (C-Ni)$_x$, where 'x' refers to the decomposition temperature. The carbon samples derived from the thermal decomposition of pristine agarose aerogel at 400 °C, 600 °C and 800 °C (labelled CA$_{400}$, CA$_{600}$ and CA$_{800}$ respectively) and the Ni-C nanoparticles prepared via thermal decomposition of pure NiAc at 400 °C, 600 °C and 800 °C (labelled (Ni-C)$_{400}$, (Ni-C)$_{600}$ and (Ni-C)$_{800}$ respectively) were used as reference samples. All the above samples were ground to a fine powder using an agate mortar and used for further characterizations.

The following techniques were used to characterize the prepared samples: (i) Thermogravimetric analysis (TGA), (ii) X-ray diffraction (XRD), (iii) Field emission scanning electron microscopy (FESEM), (iv) Energy dispersive x-ray spectroscopy (EDS), (v) Transmission electron microscopy (TEM), (vi) Multi-point BET (Brunauer-Emmett-Teller) analysis, (vii) X-ray photoelectron spectroscopy (XPS), (viii) Raman spectroscopy and (ix) Pressure composition isotherms (PCI).

To effectively engineer the distribution of nickel nanoparticles in carbon matrix (derived from agarose), a detailed study on the morphology, structure, and thermal degradation of the precursors, i.e., NiAc and agarose, is essential. So, initially the decomposition temperature of precursors i.e. NiAc, agarose and A-NiAc aerogel samples were studied using TGA (Mettler Toledo, model: Star DSC 1). The measurements were taken in $N_2$ atmosphere from 25 °C to 1000 °C at a ramping rate of 5 °C/min. XRD studies were performed on a Cu Kα radiation (λ = 1.5418 Å) Panalytical Empyrean diffractometer and the measurements were taken at 2θ range of 10°-100° with a step size of 0.01° to analyse the structure of the prepared nanocomposites. Field emission scanning electron microscope (FESEM) images using Nova Nanosem 450 were used to understand the morphology of the samples. The powder samples were spread evenly on copper tape and gold coated for 120 s prior to taking images. The images were captured at a working distance of 5 mm at an accelerating voltage of 15 kV. The elemental composition of the various samples was determined using energy-dispersive x-ray spectroscopy (EDS). Transmission electron microscopy (FEI Technai G2 12 Twin 120 kV) was used to further corroborate the results and know the distribution of Ni-C nanoparticles in the nanocomposite. The samples for TEM analysis were prepared by drop-casting a uniform suspension of nanoparticles on a 300 mesh formvar grid (Ted Pella, Inc. USA).

The surface area and pore size distribution of the various samples were determined using surface area analyser (Autosorb iQ; Quantachrome Instruments). Multi-point Brunauer-Emmett-Teller (BET) method was used to calculate the specific surface area from the nitrogen adsorption/desorption isotherms and Barett, Joyner and Halenda (BJH) adsorption plot was used to determine the pore size distribution. The samples were degassed at 300 °C for 10 hrs prior to BET measurements. To identify the changes in the chemical states of various surface elements on carbonization, X-ray photoelectron spectroscopy (XPS) (PHI 5000 Versa Prob II, FEI Inc.) was employed.

To examine the presence of adsorbed molecular hydrogen on hydrogenation and to know the structural changes in various samples, Raman spectroscopy (STR-300) was used. Data was collected between 500 cm$^{-1}$ and 2900 cm$^{-1}$ for room temperature measurements and between 500 cm$^{-1}$ and 4500 cm$^{-1}$ for hydrogenated samples. The detailed procedure for testing the hydrogenated samples is given elsewhere [11]. Finally, the samples were tested for hydrogen storage capacity in terms of weight percentage. Hydrogen sorption isotherms were obtained from a gas reaction controller (AMC, USA). Typically, 200 mg of the sample was transferred in an SS316L stainless steel sample chamber (inner diameter = 0.43 and length = 1.6 inches). Prior to the experiments, the samples were degassed at 200 °C for 12 hrs. The isotherms were obtained by dosing H$_2$ gas (99.99% purity) into the sample chamber of known volume in a step-wise fashion with sufficient equilibration between pressure dosing. LaNi$_5$ (298 K) and Cu-BTC Basolite (273 K and 77 K) obtained from Sigma Aldrich were used as reference samples for calibration. The data was collected at three different temperatures of 77 K, 273 K and 298 K as a function of hydrogen pressure (10$^{-3}$ atm to 20 atm). The calibration results are presented in the supplementary material (Fig. S2).

## 3. Results and discussion

The TGA and derivative thermogram (DTG) profile obtained for NiAc, agarose aerogel and A-NiAc aerogel are shown in Fig. 2. The DTG of NiAc revealed three decomposition stages (Fig. 2a). The first stage begins with the loss of weakly bound or adsorbed water molecules from the parent salt (Eq. 1). The major weight loss (~49%) occurred after the dehydration step with the decomposition of the intermediate acetate group within a narrow temperature range of 300 to 360 °C with the evolution of a mixture of reducing gases (CO and H$_2$) (Eq. 2). Above 360 °C, a series of side reactions took place along with the occurrence of a main reaction resulting in the formation of a metastable Ni$_3$C phase (at about 390 °C), which readily decomposes into Ni and amorphous C in the presence of an inert atmosphere (Eq. 3-6) [37].

$$Ni(CH_3COO)_2 \cdot 4H_2O \rightarrow Ni(CH_3COO)_2 + 4H_2O \tag{1}$$

$$Ni(CH_3COO)_2 \rightarrow NiCO_3 + CH_3COCH_3 \tag{2}$$

$$NiCO_3 \rightarrow NiO + CO_2 \tag{3}$$

$$CH_3COCH_3 \rightarrow CO + C_2H_6 \tag{4}$$

$$NiO + CO \rightarrow Ni + CO_2 \tag{5}$$

$$Ni_3C \rightarrow 3Ni + C \tag{6}$$

The weight derivative curve of agarose aerogel depicts two weight loss stages (Fig. 2b). The first step is associated with the volatilization of water molecules from the polysaccharide. The second weight loss (200-400 °C), with a peak temperature at 290 °C indicates the breakdown of the agarose polymeric chain and is accompanied by the release of volatile gases like $CO_2$ and CO. The carbonization of the dry porous material started after 400 °C [38]. Further, TGA characterization was also performed on A-NiAc aerogel mixed at ratio of 1:10 (Fig. 2c). It is evident that the incorporation of NiAc into the polymer raised the temperature of decomposition and the weight loss corresponding to temperature between 200 °C and 800 °C is 86%. Complete decomposition (~70%) of the samples was observed at about 400 °C with the residual weight (~30%) corresponding to metallic Ni and carbon. Based on the above analysis, three temperatures were chosen for the study i.e. 400 °C, 600 °C and 800 °C.

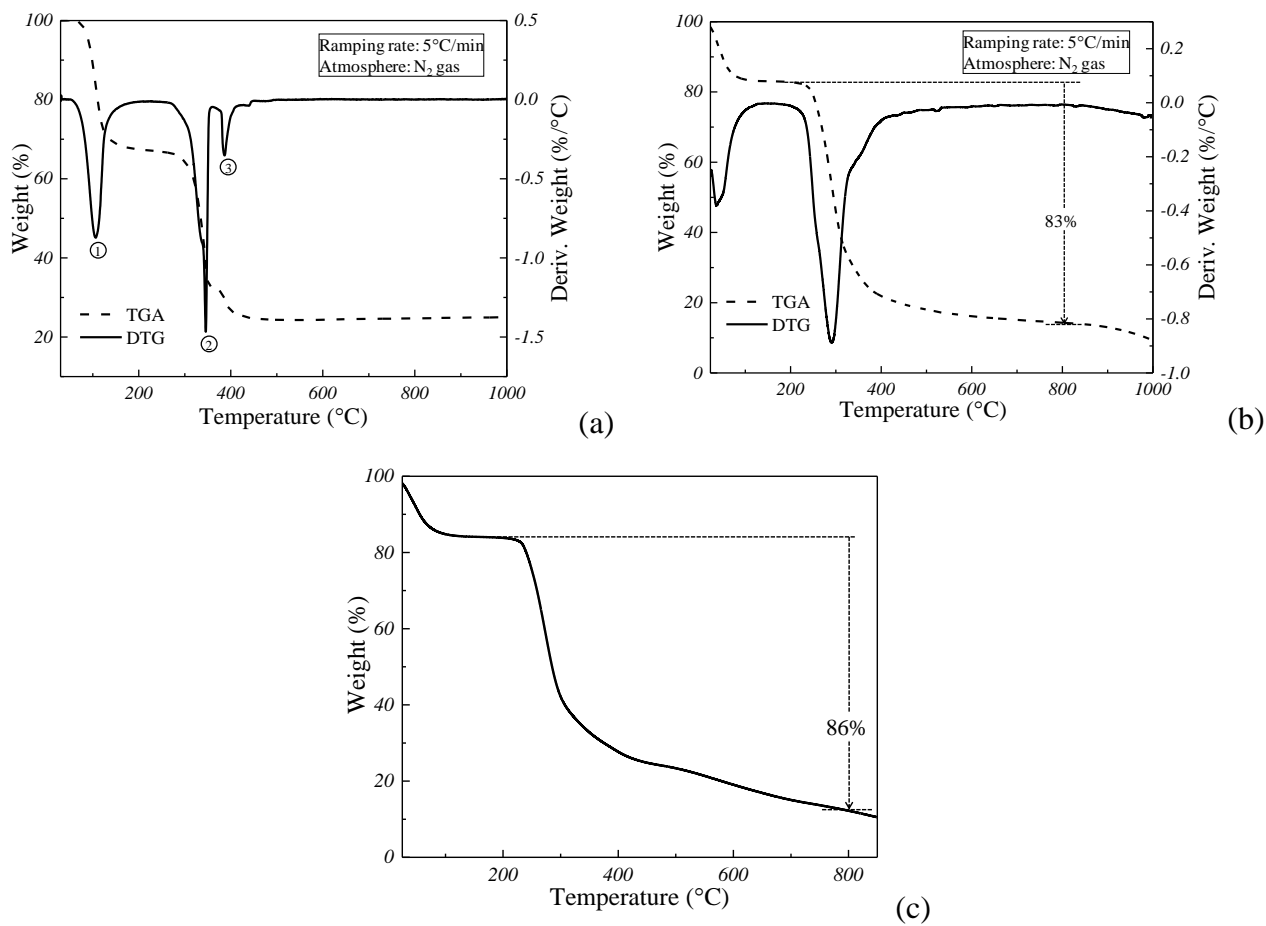

Fig. 2. TGA and DTG of (a) NiAc and (b) agarose aerogel showing pyrolysis reactions to obtain Ni-C nanoparticle from NiAc and carbon from agarose aerogel and (c) TGA profile of A-NiAc showing complete decomposition at ~400 °C.

The powder X-ray diffraction pattern of Ni-C nanoparticle revealed five characteristic peaks along with an amorphous carbon hump (Fig. 3a). The five characteristic peaks positioned at 44.5°, 51.9°, 76.4°, 92.9° and 98.5° were assigned to (111), (200), (220), (311) and (222) planes of FCC Ni [31]. A small peak at $2\theta \approx 26.4°$ which corresponds to the presence of graphitic carbon became prominent with the increase in the decomposition temperature. The diffraction pattern of $CA_{400}$, $CA_{600}$ and $CA_{800}$ samples are shown in Fig. 3b. The low intense, broad peaks shifted towards higher 2θ positions with increasing temperature and were positioned at ~23.3° (002) and ~44.0° (101) at 800 °C. The absence of sharp peaks also suggested that all the prepared CA samples were largely amorphous in nature. The interlayer spacing calculated for $CA_{800}$ at 2θ = 23°-24° was between 3.8 Å to 3.7 Å, and is marginally higher than that of graphite (3.36 Å), indicating that the structure becomes similar to graphite with increasing temperature. Furthermore, the red shift in the (002) peak validated the decrease in the interlayer spacing [39]. The increase in the diffraction peak intensity and the emergence of a new peak at 2θ ≈ 44° with the increase in the carbonization temperature, also confirmed the graphitization. X-ray diffractograms of $(C-Ni)_{400}$, $(C-Ni)_{600}$ and $(C-Ni)_{800}$ nanocomposites (Fig. 3c) showed a cumulation of the above patterns with peak positions assigned to zero valent FCC Ni crystalline phases and carbon (002) phase. Due to the reduced Ni particle size, only the (111) reflection at 2θ = 44.5° could be resolved as a weak peak for samples carbonized at 400 °C. The crystallite size of Ni-C nanoparticle derived from the (111) reflection using the Scherrer equation for $(C-Ni)_{400}$, $(C-Ni)_{600}$ and $(C-Ni)_{800}$ samples were 1.2 nm, 2.8 nm and 24.8 nm respectively, whereas those of free Ni-C nanoparticles were 33.6 nm (400 °C), 39.4 nm (600 °C) and 46.8 nm (800 °C). These findings imply that the carbon matrix not only assists in the homogenous distribution of the nanoparticles (Fig. 4) but also aids in controlling the particle size. The reduction in metal nanoparticle size when confined in a porous matrix is in line with the experimental findings of other researchers [22,23]. The results also confirmed that no diffraction peaks corresponding to other phases were present (i.e. NiO).

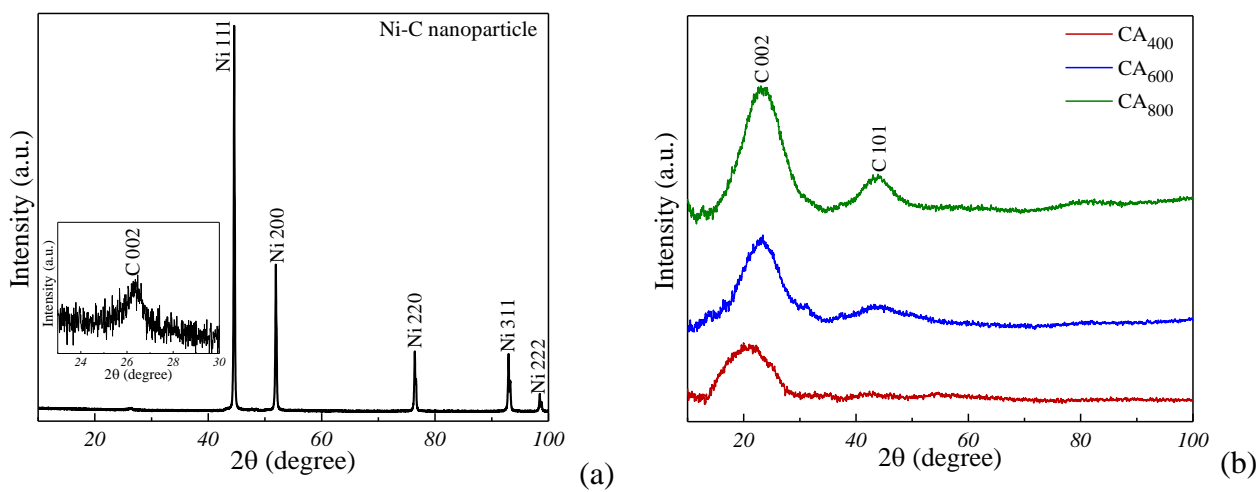

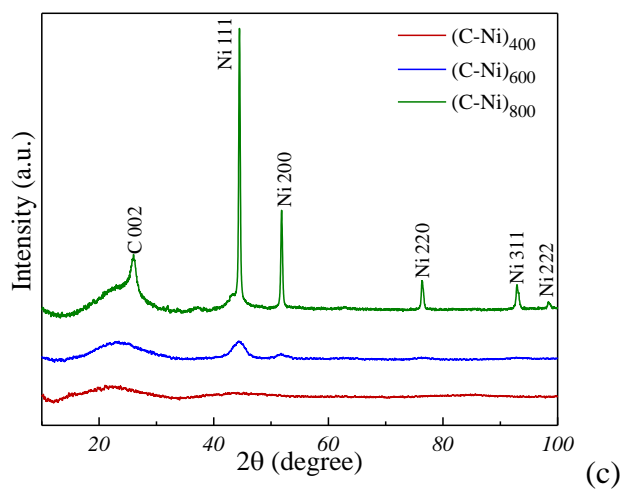

Fig. 3. Powder XRD pattern of as-prepared: (a) Ni-C nanoparticle with inset highlighting the carbon hump at 2θ = ~26.4°, (b) $CA_{400}$, $CA_{600}$ and $CA_{800}$ and (c) $(C-Ni)_{400}$, $(C-Ni)_{600}$ and $(C-Ni)_{800}$ nanocomposite. With an increase in decomposition temperature, the broad diffraction peak assigned to graphite C (002) sharpens and indicates a rise in graphitization.

To get insights into the morphological evolution of the heat-treated samples, FESEM and TEM-based analysis was performed and is presented in Fig. 4. The product obtained after the decomposition of NiAc appeared as spherical aggregates (Fig. 4a1) with a mean particle size of ~ 155 nm as shown in the histogram (Fig. 4a4). A closer look at the TEM images revealed irregular and nearly spherical nickel nanoparticles surrounded by a thin carbon ring of ~ 12 nm thicknesses (Fig. 4a2). The carbon capsule encasing the metal nanoparticles protects them from converting to NiO when exposed to air, which could otherwise potentially hinder the storage capacity. The results also comply with the XRD results (absence of NiO peaks) (Fig. 4a3). The elemental percentages as determined using EDS for C, Ni and O were 4.6, 95 and 0.4 wt.% respectively with no additional impurities (Table. T1 of supplementary material). The particle morphology, size and distribution were more pronounced after using the carbon matrix (Fig. 4b2, c2 and d2). Results indicate that the mean particle diameter decreased (Fig. 4b4, c4 and d4) and the particles were homogeneously spread on the 2D sheet-like carbon with a highly narrow size distribution. It is believed that the agarose hydrogel provides a matrix for the fine distribution of the precursor material which upon freeze-drying and carbonization produces a homogeneous dispersion of Ni-C nanoparticles in carbon matrix. The carbon matrix present around the Ni-C nanoparticle also inhibits the growth of the particles giving rise to smaller and stable nanoparticles. Such nanoparticles prepared were stable at room temperature due to the thin carbon layer surrounding them. Additionally, the nucleation and growth of the Ni-C nanoparticle in the matrix can be controlled by changing the heat treatment temperature. The SAED rings were indexed to stable FCC crystalline Ni phase, suggesting no change in the crystalline nature of the samples upon employing the carbon matrix (Fig. 4b3, c3 and d3).

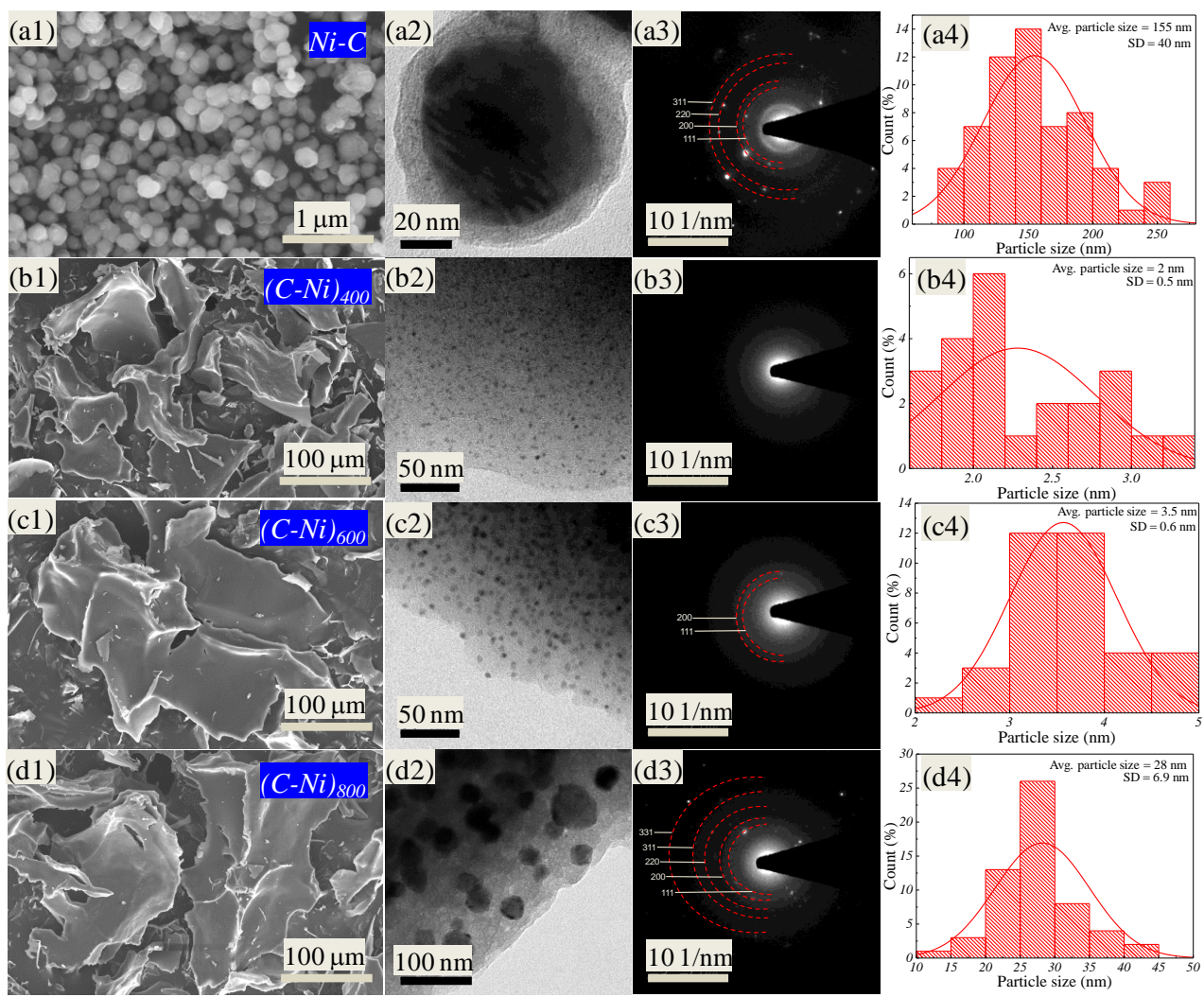

Fig. 4. FESEM and TEM images of: (a1 and a2) Ni-C nanoparticle, (b1 and b2) (C-Ni)$_{400}$, (c1 and c2) (C-Ni)$_{600}$ and (d1 and d2) (C-Ni)$_{800}$ along with the respective SAED pattern (a3, b3, c3 and d3) and particle size histogram of corresponding Ni-C nanoparticles (a4, b4, c4 and d4). The carbonized agarose matrix as seen in b2, c2 and d2 appears in lighter contrast when compared to the electron-dense Ni nanoparticles.

The average size of the nanoparticles increased with the annealing temperature. The average particle size and the standard deviation of the Ni-C nanoparticles embedded in the carbon substrate are tabulated in Table 1. The consolidated EDS analysis data of the Ni-C (no carbon matrix) and C-Ni (with carbon matrix) nanocomposites are provided in the supplementary material (Table. T1).

Table 1. Comparative study of the average particle size along with standard deviation of the Ni-C nanoparticles in C-Ni nanocomposite obtained at various heat treatment temperatures.

| Sample | Avg. particle size (nm) |
| --- | --- |
| (C-Ni)$_{400}$ | 2.0 ± 0.5 |
| (C-Ni)$_{600}$ | 3.5 ± 0.6 |
| (C-Ni)$_{800}$ | 28.0 ± 6.9 |

Raman spectroscopy was performed to ascertain the nature of the carbon shell surrounding the Ni core in the Ni-C nanoparticle as well as that of the carbon in various CA and C-Ni samples. Fig. 5a presents the Raman spectra of NiAc carbonized at various temperatures. It was characterized by the presence of a D peak (disordered carbon) at ~1349 cm$^{-1}$ and a G peak (graphitic carbon) at ~1582 cm$^{-1}$. The consistent increase in the intensity of the graphitic carbon peak relative to that of the disordered carbon and the emergence of a 2D peak upon increasing the decomposition temperature, suggests a gradual progress from amorphous carbon to highly ordered and graphitic carbon around the Ni core. The $I_D/I_G$ ratio also revealed a strong dependence of the sample's structural quality with the annealing temperature i.e. the samples heat treated at lower temperatures were more defective with poor crystallinity. The Raman spectra of CA and C-Ni samples also exhibited two prominent peaks corresponding to the D and G bands (Fig. 5b, c and S7). However, in contrast to the Raman spectra of Ni-C nanoparticles, the $I_D/I_G$ ratio increased from 0.63 to 0.91 for CA and from 0.64 to 0.95 for C-Ni nanocomposite when heat treated from 400 °C to 800 °C. This indicates the existence of more distortion and disorder in the carbon structure at higher temperatures, probably due to the disruption of graphitic domains [40]. The steady rise in the intensity of the D peak further substantiates that such defects may originate due to sp$^3$ defective carbon [41]. Additionally, with an increase in the annealing temperature, a red shift in the G-band and a blue shift in the D-band were also observed.

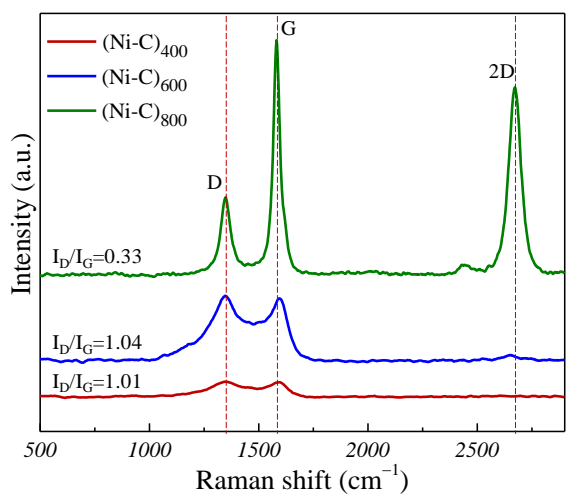
(a)

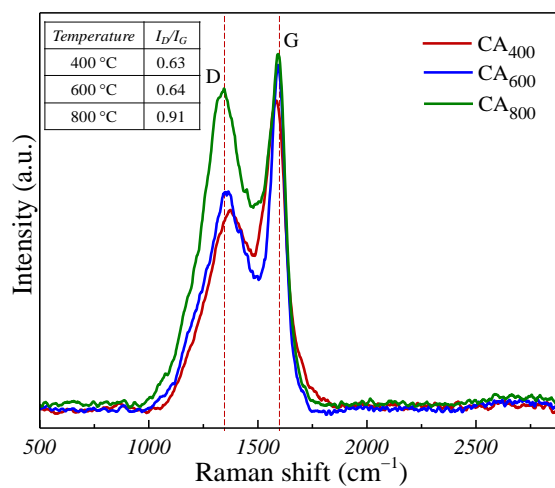
(b)

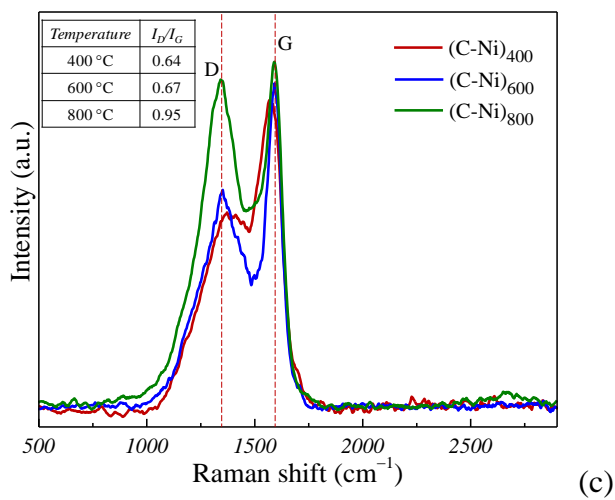

Fig. 5. Raman spectra of: (a) (Ni-C)$_{400}$, (Ni-C)$_{600}$ and (Ni-C)$_{800}$ nanoparticles showing the emergence of 2D peak with increase in heat treatment temperature, (b) CA$_{400}$, CA$_{600}$ and CA$_{800}$ and (c) (C-Ni)$_{400}$, (C-Ni)$_{600}$ and (C-Ni)$_{800}$ nanocomposite showing an enhancement in carbon defect states ($I_D/I_G$) with increase in temperature.

XPS analysis shown in Fig. 6 was performed to probe the surface oxidation states of C and Ni, and thereby study the evolution of their chemical states in Ni-C nanoparticles when confined in a carbon matrix. The survey spectra of Ni-C nanoparticles captured over a broad energy range showed a strong carbon signal at 284.7 eV, confirming the existence of a carbon layer around the metal core. Results reveal that the +2 oxidation state of NiAc precursor was reduced to zero-valent Ni upon heat treatment. A detailed examination of the high-resolution Ni 2p spectrum confirmed the presence of both Ni$^{2+}$ and Ni$^{0}$ (Fig. 6b). The Ni 2p$_{3/2}$ and Ni 2p$_{1/2}$ peaks which represent the oxidation state of Ni were centered at 855.6 eV and 873.6 eV with the respective satellite peaks at 861.0 eV and 879.9 eV and are presumed to come from the C-Ni interface [42]. Ni$^{0}$ peaks were positioned at 852.9 eV and 870.6 eV and arose from the metallic nickel core. The high-resolution C 1s spectrum was deconvoluted into two prominent peaks centered at 284.7 eV and 285.9 eV and were assigned to C=C and C-C. But after the introduction of oxygen rich carbon matrix, a new peak can be deconvoluted at 288.1 eV and corresponds to the presence of C-O/C=O functional group. These oxygen-rich functional groups present in the carbonized agarose facilitate in the effective passivation of confined metal nanoparticles. Compared to the unsupported Ni-C nanoparticle, the peak intensity of Ni$^{0}$ in the Ni 2p spectrum of C-Ni nanocomposite was reduced and showed a positive shift (~0.3 eV) in the binding energy, suggesting a change in the electronic structure of Ni after the introduction of carbon matrix. The positive shift further indicates a reduction in the electron density around Ni giving rise to empty d-orbitals, which in turn helps in improving its binding ability with hydrogen, ultimately enhancing the hydrogen storage performance [43,44].

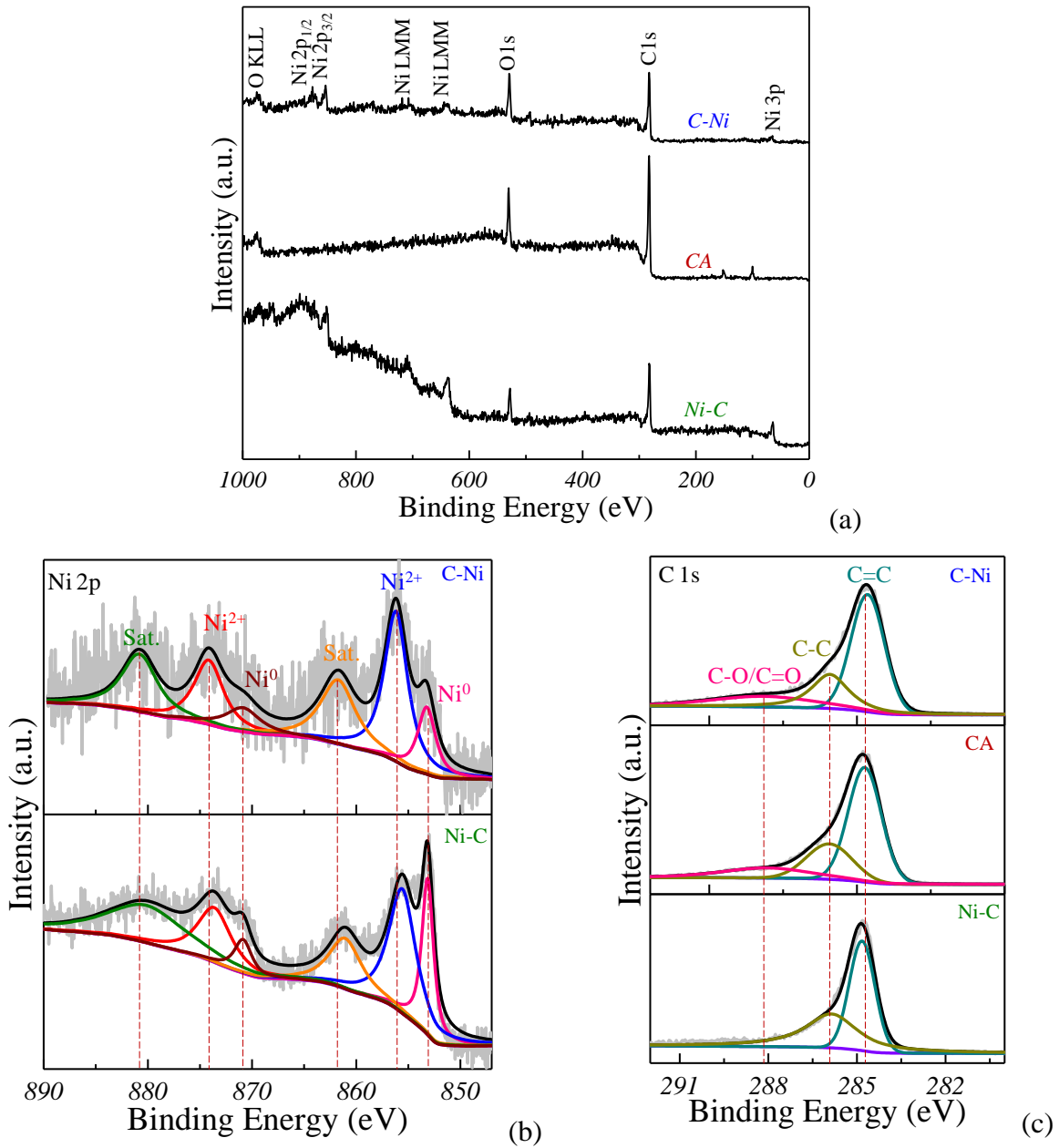

Fig. 6. XPS survey spectra of: (a) (Ni-C)$_{800}$ nanoparticle, (CA)$_{800}$ and (C-Ni)$_{800}$ nanocomposite showing the presence of C, O and Ni elements, (b) Comparative high-resolution Ni 2p spectra of Ni-C nanoparticle and C-Ni nanocomposite illustrating the coexistence of Ni$^{2+}$ and Ni$^{0}$ and (c) Comparative high-resolution C 1s spectra of Ni-C nanoparticle, CA and C-Ni samples.

The prepared samples were examined for hydrogen storage capacity. Fig. 7 shows the room temperature (298 K) absorption isotherms measured as a function of H$_2$ pressure. The absorption capacity of all the Ni-C samples was found to increase linearly with an increase in pressure, suggesting that the carbon layer around Ni is gas-permeable to allow effective adsorption of hydrogen. Among the synthesized samples, (Ni-C)$_{400}$ sample showed a maximum hydrogen uptake of 0.079 wt.% (Fig. 7a). As there is no saturation observed, it could be possible that the storage capacity can be further enhanced by increasing the pressure. The hydrogen storage capacity of all the prepared CA samples increased with an increase in the heat treatment temperature (Fig. 7b). This behaviour can be ascribed to the increase in the carbon weight percentage and the subsequent rise in the surface area with heat treatment temperature, which further aids in providing more surface sites for the attachment of hydrogen, thereby improving the storage capacity [45]. Table. T1 and T2 of the supplementary material gives the carbon weight percentage and the surface area of CA samples heat treated at 400 °C, 600 °C and 800 °C obtained from EDS and BET respectively. $CA_{800}$ having higher surface area of 362 m$^2$/g showed a maximum hydrogen storage of 5.7 wt.%, 0.87 wt.% and 0.11 wt.% at 77 K, 273 K and 298 K respectively at 20 atm $H_2$ pressure (Supplementary material, Fig S3).

Fig. 7c shows the hydrogen absorption curve of C-Ni nanocomposites. It was observed that the hydrogen storage capacity of C-Ni nanocomposites at room temperature was significantly more when compared to that of Ni-C and CA samples. It is to be noted that the hydrogen storage at room temperature is primarily attributed to the spillover mechanism as the formation of nickel hydride is thermodynamically favourable only at very high pressure i.e., >5000 bar [46,47]. Despite the decrease in surface area, the (C-Ni)$_{400}$ nanocomposite showed a maximum hydrogen uptake of 0.73 wt.% at 298 K, which is 6.5 times greater than that of the pure CA sample. Fig. 7d shows the cyclic stability of the (C-Ni)$_{400}$ nanocomposite for three absorption-desorption cycles. It can be seen that the C-Ni nanocomposite showed very good cyclability with negligible hysteresis. The weak adsorption strength facilitates the easy recombination and desorption of the adsorbed species from the support at ambient pressure and room temperature. Fig. 7e compares the maximum hydrogen storage in the prepared Ni-C, CA and C-Ni nanocomposite at 298 K. Results indicate that by nanostructuring/engineering, a remarkable improvement in the hydrogen storage capacity can be achieved in the materials.

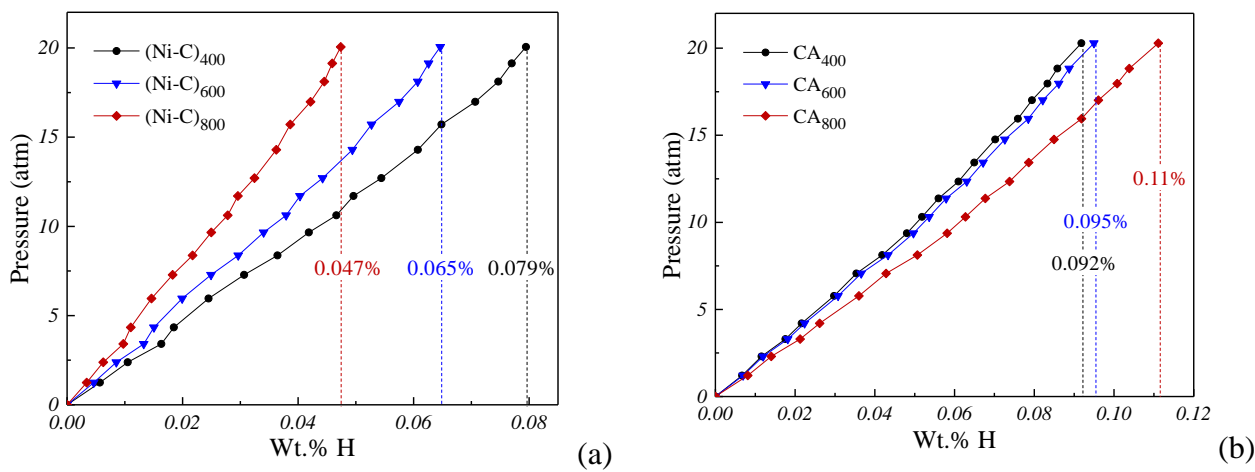

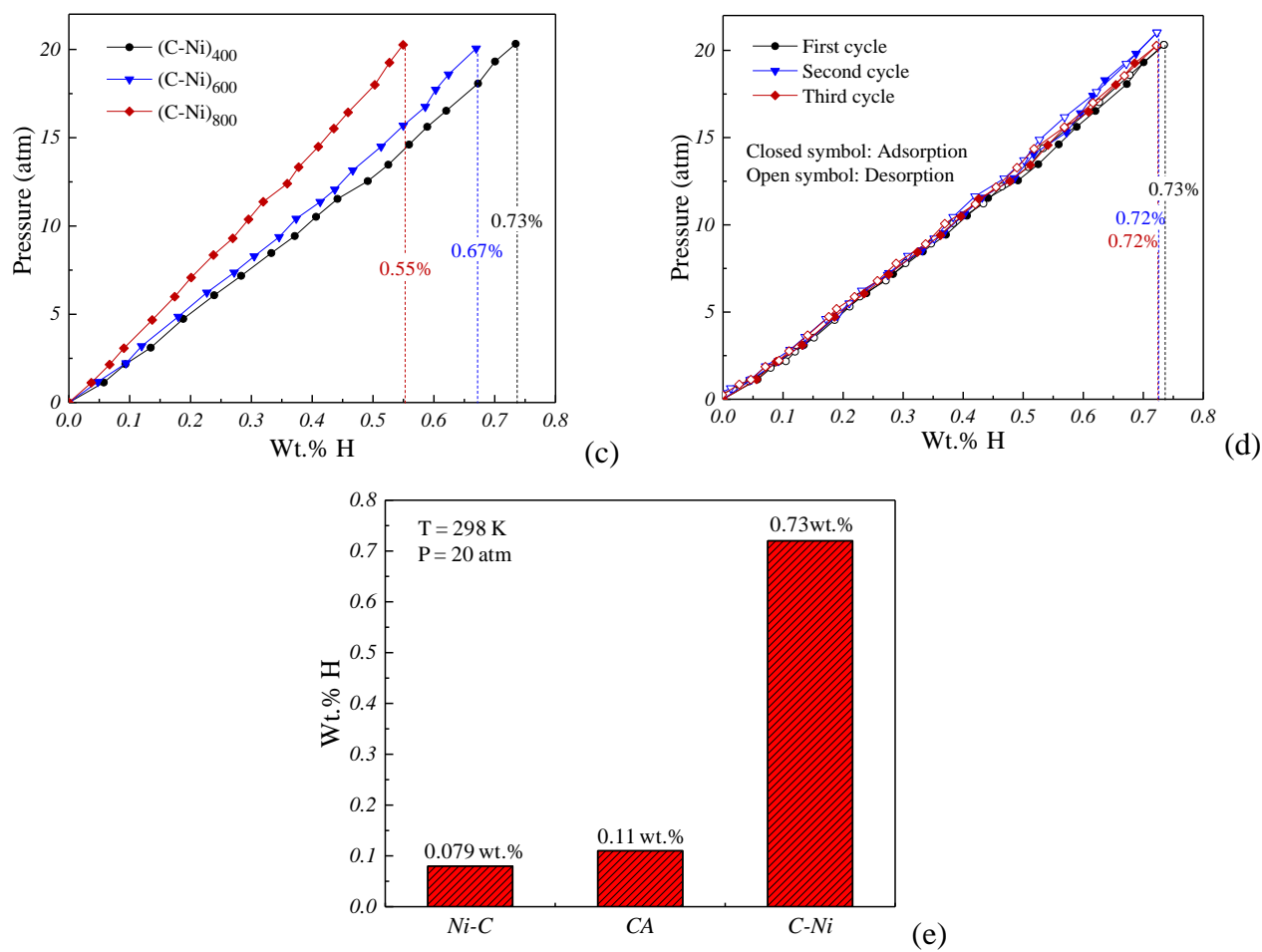

Fig. 7. Hydrogen sorption isotherm measured at 298 K for: (a) Ni-C, (b) CA and (c) C-Ni nanocomposite at 20 atm, (d) Hydrogen adsorption and desorption cycles at 20 atm and 298 K and (e) Comparison plot showing the maximum hydrogen uptake of Ni-C, CA and C-Ni nanocomposite.

The increase in hydrogen storage capacity in the C-Ni nanocomposites was a direct consequence of the presence of nickel nanoparticles, which invokes the spillover and the presence of high surface area porous carbon structure which acts as additional absorption sites for the dissociated hydrogen atoms. Further, the particle size of Ni-C also dictates the hydrogen storage capacity i.e., smaller the particle size, more efficient the catalytic activity of the Ni nanoparticles aiding the spillover mechanism. The schematic of the spillover mechanism is depicted in Fig. 8.

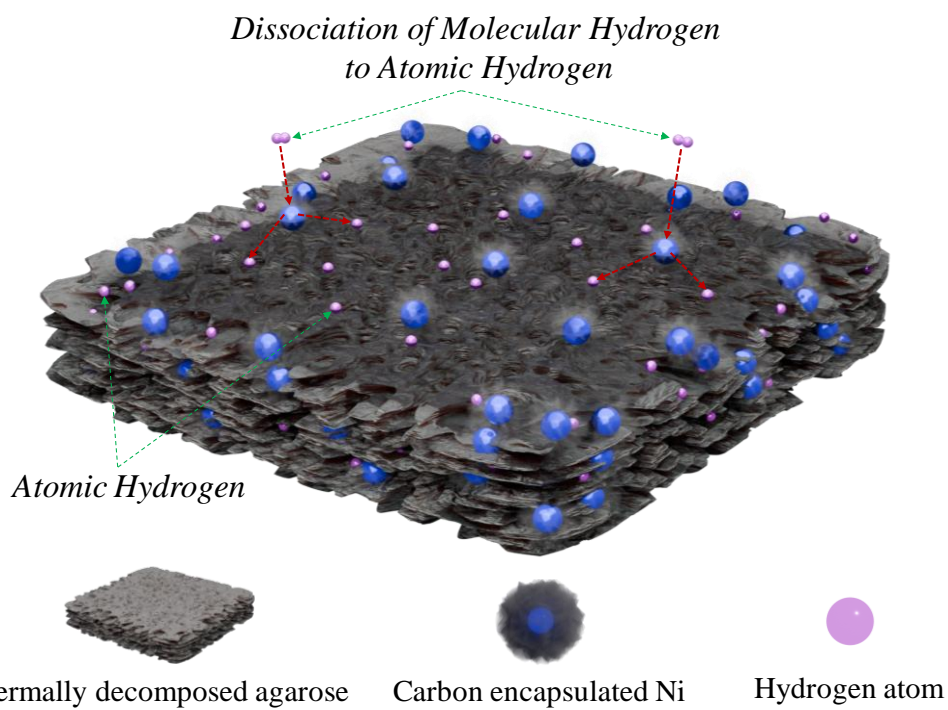

Fig. 8. Schematic representation of the spillover mechanism demonstrating the dissociation of free molecular hydrogen into atomic counterparts in the presence of a metal catalyst (Ni) and the subsequent surface diffusion of the H atoms.

Raman spectroscopy was used to demonstrate the presence of molecular hydrogen physisorbed on the (C-Ni)$_{400}$ sample at 77 K and the results are shown in Fig. 9. The unhydrogenated sample (T = 298 K and P = 1 atm) showed the presence of only D and G peak at ~1340 cm$^{-1}$ and ~1583 cm$^{-1}$. However, when the sample was exposed to an H$_2$ pressure of 20 atm at 273 K, two strong peaks corresponding to rotational and vibrational modes of molecular hydrogen emerged at ~586 cm$^{-1}$ and ~4153 cm$^{-1}$. These peaks gradually became more pronounced as the temperature was further reduced to 77 K. Another notable observation was the absence of Raman active C-H stretching mode between 2800-3300 cm$^{-1}$ which eliminates the possibility of chemisorption at these temperatures [48]. These peaks tend to disappear when the temperature was raised above room temperature.

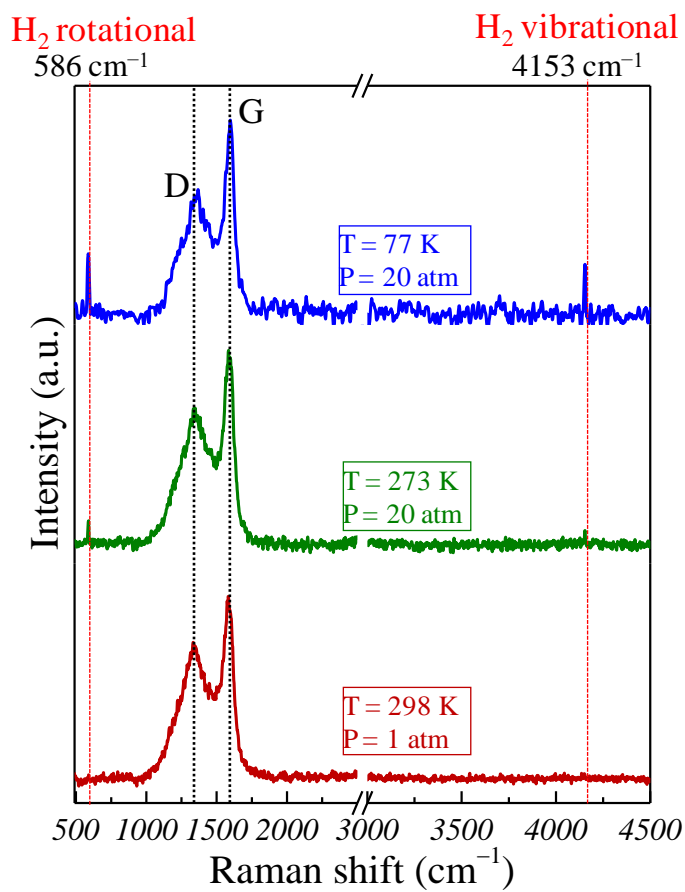

Fig. 9. Raman spectra of the hydrogenated (C-Ni)$_{400}$ nanocomposite obtained between 500 and 4500 cm$^{-1}$ at: (a) 298 K, (b) 273 K, and (c) 77 K.

## 4. Summary and conclusions

In this work, we report a simple, industrially scalable and cost-effective approach to synthesize carbon-based nanocomposites containing catalytic Ni nanoparticles for hydrogen storage application. The metallic nanoparticles obtained thus were stable and resilient to atmospheric oxidation due to the thin carbon layer encapsulating them. Additionally, the use of biomass-derived carbon matrix aided in four ways: (i) helped in the homogenous distribution of the metallic nanoparticles without agglomeration by cross-linking with the metal ions, (ii) effectively helped in controlling the Ni-C particle size, further assisting in efficient catalytical activity, (iii) served as potential adsorption sites for molecular hydrogen, (iii) acted as secondary absorption sites for atomic hydrogen after the spillover mechanism. Nickel was used as the transition metal to trigger the spillover mechanism wherein the molecular hydrogen dissociates to its atomic counterparts and gets adsorbed on the surface of the carbon matrix. Thus, the uniform dispersion of catalytic nanoparticles embedded in the carbon matrix favoured the enhancement of hydrogen storage capacity at near-ambient conditions. It was shown that by efficient nanostructuring, the (C-Ni)$_{400}$ nanocomposite exhibited a good reversible storage capacity of 0.73 wt.% (against 0.11 wt.% for pure carbon sample) at room temperature (298 K) and 20 atm $H_2$ pressure. The study further concludes that a significant storage capacity can be achieved by effectively optimizing the metal content, size and textural properties of the carbon matrix. These results are far superior to the reported storage capacities of other noble metal (Pd, Pt) based nanocomposites.

**Acknowledgments**

We thank Mr. Shilankar (Raman and Photoluminescence Laboratory, I.I.T Kanpur) and Mr. Jai Singh (Advanced Imaging Center, I.I.T Kanpur) for facilitating Raman spectroscopy and TEM. The study was financially supported by the Department of Science and Technology (DST) via grant nos. DST/TMD/MECSP/2K17/14 (C) and DST/TMD/IC-MAP/2K20/02 (C). A Flamina would like to thank the Ministry of Human Resource Development (MHRD) for the scholarship via the Prime Minister's Research Fellowship (PMRF) Scheme.

**References**

[1] R. Moliner, M. J. Lazaro and I. Suelves. Analysis of the strategies for bridging the gap towards the Hydrogen Economy. *International Journal of Hydrogen Energy* **41**(43), 19500-19508 (2016). http://dx.doi.org/10.1016/j.ijhydene.2016.06.202

[2] Jason Graetz. New approaches to hydrogen storage. *Chem. Soc. Rev.,* **38**, 73–82 (2009). https://doi.org/10.1039/B718842K

[3] Jun Yang, Andrea Sudik, Christopher Wolverton and Donald J. Siegel. High capacity hydrogen storage materials: attributes for automotive applications and techniques for materials discovery. *Chem. Soc. Rev.,* **39**, 656–675 (2010). https://doi.org/10.1039/B802882F

[4] Puru Jena. Materials for Hydrogen Storage: Past, Present, and Future. *J. Phys. Chem. Lett.,* 2, 206–211 (2011). DOI: 10.1021/jz1015372.

[5] Yongde Xia, Zhuxian Yang and Yanqiu Zhu. Porous carbon-based materials for hydrogen storage: advancement and challenges. *J. Mater. Chem. A* **1**, 9365-9381 (2013). https://doi.org/10.1039/C3TA10583K.

[6] US Department of Energy (2023). DOE Technical Targets for On-board Hydrogen Storage for Light-Duty Vehicles. Available online at: https://www.energy.gov/eere/fuelcells/doe-technical-targets-onboard-hydrogen-storage-light-duty-vehicles

[7] Alexander Schoedel, Zhe Ji and Omar M. Yaghi. The role of metal-organic frameworks in a carbon-neutral energy cycle. *Nat Energy* **1**, 16034 (2016). https://doi.org/10.1038/nenergy.2016.34

[8] Lifeng Wang and Ralph T. Yang. New sorbents for hydrogen storage by hydrogen spillover – a review. *Energy Environ. Sci.,* **1**: 268-279 (2008). https://doi.org/10.1039/B807957A.

[9] Fenil J. Desai, Md Nizam Uddin, Muhammad M. Rahman and Ramazan Asmatulu. A critical review on improving hydrogen storage properties of metal hydride via nanostructuring and integrating carbonaceous materials. *IJHE* **48**: 29256-29294 (2023). https://doi.org/10.1016/j.ijhydene.2023.04.029

[10] Li Ren, Yinghui Li, Ning Zhang, Zi Li, Xi Lin, Wen Zhu, Chong Lu, Wenjiang Ding and Jianxin Zou. Nanostructuring of Mg-Based Hydrogen Storage Materials: Recent Advances for Promoting Key Applications. *Nano-Micro Lett.* **15**:93 (2023). https://doi.org/10.1007/s40820-023-01041-5

[11] A. Flamina, R.M. Raghavendra, Anshul Gupta and Anandh Subramaniam. Hydrogen storage in Nickel dispersed boron doped reduced graphene oxide. *Applied Surface Science Advances* **13**: 100371 (2023). https://doi.org/10.1016/j.apsadv.2023.100371

[12] Rupali Nagar, Sumita Srivastava, Sterlin Leo Hudson, Sandra L. Amaya, Ashish Tanna, Meenu Sharma, Ramesh Achayalingam, Sanjiv Sonkaria, Varsha Khare and Sesha S. Srinivasan. Recent developments in state-of-the-art hydrogen energy technologies – Review of hydrogen storage materials. *Solar Compass* **5**: 100033 (2023). https://doi.org/10.1016/j.solcom.2023.100033.

[13] Suboohi Shervani, Puspal Mukherjee, Anshul Gupta, Gargi Mishra, Kavya Illath, T.G. Ajithkumar, Sri Sivakumar, Pratik Sen, Kantesh Balani and Anandh Subramaniam, Multi-mode hydrogen storage in nanocontainers. *IJHE* **42**: 24256-24262 (2017), 10.1016/j.ijhydene.2017.07.233.

[14] Flamina Amaladasse, Anshul Gupta, Suboohi Shervani, Sri Sivakumar, Kantesh Balani and Anandh Subramaniam, Enhanced reversible hydrogen storage in palladium hollow spheres, *Part. Sci. Technol.* **39**: 617–623 (2020). https:// doi.org/10.1080/02726351.2020.1776433.

[15] Anshul Gupta, Suboohi Shervani, Flamina Amaladasse, Sri Sivakumar, Kantesh Balani and Anandh Subramaniam, Enhanced reversible hydrogen storage in nickel nano hollow spheres. *IJHE,* **44**: 22032-22038 (2019). 10.1016/j.ijhydene.2019.06.090

[16] Anshul Gupta, Suboohi Shervani, Pooja Rani, Sri Sivakumar, Kantesh Balani and Anandh Subramaniam, Hybrid hollow structures for hydrogen storage. *IJHE,* **45**: 24076-24082 (2020). 10.1016/j.ijhydene.2019.03.273

[17] Jacob Burress, Michael Kraus, Matt Beckner, Raina Cepel, Galen Suppes, Carlos Wexler and Peter Pfeifer. Hydrogen storage in engineered carbon nanospaces. *Nanotechnology* **20**: 204026 (2009). http://dx.doi.org/10.1088/0957-4484/20/20/204026

[18] Dong Ju Han, Ki Ryuk Bang, Hyun Cho and Eun Seon Cho. Effect of carbon nanoscaffolds on hydrogen storage performance of magnesium hydride. *Korean J. Chem. Eng.,* **37**(8): 1306-1316 (2020). https://doi.org/10.1007/s11814-020-0630-2

[19] Jingjing Zhang, Bing Zhang, Xiubo Xie, Cui Ni, Chuanxin Hou, Xueqin Sun, Xiaoyang Yang, Yuping Zhang, Hideo Kimura and Wei Du. Recent advances in the nanoconfinement of Mg-related hydrogen storage materials: A minor review. *International Journal of Minerals, Metallurgy and Materials* **30**: 14–24 (2023). https://doi.org/10.1007/s12613-022-2519-z

[20] Yongfeng Liu, Wenxuan Zhang, Xin Zhang, Limei Yang, Zhenguo Huang, Fang Fang, Wenping Sun, Mingxia Gao and Hongge Pan. Nanostructured light metal hydride: Fabrication strategies and hydrogen storage performance. *Renewable and Sustainable Energy Reviews* **184**: 113560 (2023). https://doi.org/10.1016/j.rser.2023.113560

[21] Thomas K. Nielsen, Flemming Besenbacher and Torben R. Jensen. Nanoconfined hydrides for energy storage. *Nanoscale* **3**: 2086-2098 (2011). https://doi.org/10.1039/C0NR00725K.

[22] Petra E. de Jongh, Rudy W. P. Wagemans, Tamara M. Eggenhuisen, Bibi S. Dauvillier, Paul B. Radstake, Johannes. D. Meeldijk, John W. Geus, and Krijn P. de Jong. The Preparation of Carbon-Supported Magnesium Nanoparticles using Melt Infiltration. *Chem. Mater.* **19**(24): 6052–6057 (2007). https://doi.org/10.1021/cm702205v

[23] Shu Zhang, Adam F Gross, Sky L Van Atta, Maribel Lopez, Ping Liu, Channing C Ahn, John J Vajo and Craig M Jensen. The synthesis and hydrogen storage properties of a $MgH_2$ incorporated carbon aerogel scaffold. *Nanotechnology* **20**: 204027 (2009). http://dx.doi.org/10.1088/0957-4484/20/20/204027.

[24] Emmanuel Boateng and Aicheng Chen. Recent advances in nanomaterial-based solid-state hydrogen storage. *Materials Today Advances* **6**:100022 (2020). https://doi.org/10.1016/j.mtadv.2019.100022

[25] Darryl S. Pyle, E. MacA. Gray and C. J. Webb. Hydrogen storage in carbon nanostructures via spillover. *IJHE* **41**:19098-19113 (2016) https://doi.org/10.1016/j.ijhydene.2016.08.061

[26] Lifeng Wang and Ralph T. Yang Hydrogen Storage on Carbon-Based Adsorbents and Storage at Ambient Temperature by Hydrogen Spillover. *Catalysis Reviews: Science and Engineering*, 52:411–461 (2010). https://doi.org/10.1080/01614940.2010.520265

[27] M. Zielinski, R.Wojcieszak, S. Monteverdi, M. Mercy and M. M. Bettahar. Hydrogen storage in nickel catalysts supported on activated carbon. *IJHE* **32**: 1024-1032 (2007) https://doi.org/10.1016/j.ijhydene.2006.07.004.

[28] Minglong Zhong, Zhibing Fu, Rui Mi, Xichuan Liu, Xiaojia Li, Lei Yuan, Wei Huang, Xi Yang, Yongjian Tang, and Chaoyang Wang. Fabrication of Pt-doped carbon aerogels for hydrogen storage by radiation method. *International Journal of Hydrogen Energy* **43**:19174-19181 (2018). https://doi.org/10.1016/j.ijhydene.2018.08.169.

[29] Natthaporn Thaweelap, Praphatsorn Plerdsranoy, Yingyot Poo-arporn, Patcharaporn Khajondetchairit, Suwit Suthirakun, Ittipon Fongkaew, Pussana Hirunsit, Narong Chanlek, Oliver Utke, Autchara Pangon and Rapee Utke. Ni-doped activated carbon nanofibers for storing hydrogen at ambient temperature: Experiments and computations. *Fuel* **288**: 119608 (2021). https://doi.org/10.1016/j.fuel.2020.119608.

[30] Chen-Chia Huang, Yi-Hua Li, Yen-Wen Wang, and Chien-Hung Chen. Hydrogen storage in cobalt-embedded ordered mesoporous carbon. *International Journal of Hydrogen Energy* **38**:3994-4002 (2013). http://dx.doi.org/10.1016/j.ijhydene.2013.01.081

[31] Swati V. Pol, Vilas G. Pol, Israel Felner, and Aharon Gedanken. The Thermal Decomposition of Three Magnetic Acetates at Their Autogenic Pressure Yields Different Products. Why? *Eur. J. Inorg. Chem.*, 2089–2096 (2007). https://doi.org/10.1002/ejic.200700146.

[32] Thomas O. M. Samuels, Alex W. Robertson, Heeyeon Kim, Mauro Pasta and Jamie H. Warner. Three dimensional hybrid multi-layered graphene-CNT catalyst supports via rapid thermal annealing of nickel acetate. *J. Mater. Chem. A* **5**,10457 (2017). DOI:10.1039/c7ta01852e.

[33] BoYou, PeiqunYin, Junli Zhang, Daping He, Gaoli Chen, Fei Kang, Huiqiao Wang, Zhaoxiang Deng and Yadong. Hydrogel-derived non-precious electrocatalysts for efficient oxygen reduction. *Sci Rep* **5**, 11739 (2015). https://doi.org/10.1038/srep11739

[34] Valentina Tozzini and Vittorio Pellegrini. Prospects for hydrogen storage in graphene. *Phys. Chem. Chem. Phys.,* **15**, 80 (2013). DOI: 10.1039/c2cp42538f.

[35] Hamid Ghorbani Shiraz, Omid Tavakoli. Investigation of graphene-based systems for hydrogen storage. *Renewable and Sustainable Energy Reviews* **74**:104-109 (2017). https://doi.org/10.1016/j.rser.2017.02.052.

[36] Ismael Gonzalez, Juan C. De Jesus, Edgar Cañizales, Blas Delgado and Caribay Urbina. Comparison of the Surface State of Ni Nanoparticles Used for Methane Catalytic Decomposition. *J. Phys. Chem. C*, **116**, 21577−21587 (2012). dx.doi.org/10.1021/jp302372r.

[37] Andrew K. Galwey, Samuel G. Mckee, Thomas R.B. Mitchell, Michael E. Brown and Alison F. Bean. A kinetic and mechanistic study of the thermal decomposition of nickel acetate. *Reactivity of Solids*, **6**: 173-186 (1988). https://doi.org/10.1016/0168-7336(88)80059-7.

[38] Kaijun Xie, Xin Liu, Haolin Li, Long Fang, Kai Xia, Dongjiang Yang, Yihui Zou and Xiaodong Zhang. Heteroatom tuning in agarose derived carbon aerogel forenhanced potassium ion multiple energy storage. *Carbon Energy* e427 (2023). https://doi.org/10.1002/cey2.427.

[39] Guanggui Cheng, Lingjian Dong, Lakhinder Kamboj, Tushar Khosla, Xiaodong Wang, Zhongqiang Zhang, Liqiang Guo, Noshir Pesika, and Jianning Ding. Hydrothermal Synthesis of Monodisperse Hard Carbon Spheres and Their Water-Based Lubrication. *Tribol Lett* **65**:141 (2017). DOI: 10.1007/s11249-017-0923-8.

[40] L. Scott Blankenship, Norah Balahmar, and Robert Mokaya. Oxygen-rich microporous carbons with exceptional hydrogen storage capacity. *Nature Communications* **8**: 1545 (2017). https://doi.org/10.1038/s41467-017-01633-x.

[41] Vikrant Sahu, Ram Bhagat Marichi, Gurmeet Singh, and Raj Kishore Sharma. Multifunctional, Self-Activating Oxygen-Rich Holey Carbon Monolith Derived from Agarose Biopolymer. *ACS Sustainable Chem. Eng.* **5**:8747-875 (2017). http://dx.doi.org/10.1021/acssuschemeng.7b01543.

[42] Juan C. De Jesús, Salvador García and Daniela Dorante. Size tunable carbon-encapsulated nickel nanoparticles synthesized by pyrolysis of nickel acetate tetrahydrate. *Journal of Analytical and Applied Pyrolysis* **130**: 332–343 (2018). https://doi.org/10.1016/j.jaap.2017.12.013.

[43] Jaerim Kim, Hyeonjung Jung, Sang-Mun Jung, Jinwoo Hwang, Dong Yeong Kim, Noho Lee, Kyu-Su Kim, Hyunah Kwon, Yong-Tae Kim, Jeong Woo Han, and Jong Kyu Kim. Tailoring Binding Abilities by Incorporating Oxophilic Transition Metals on 3D Nanostructured Ni Arrays for Accelerated Alkaline Hydrogen Evolution Reaction. *J. Am. Chem. Soc.*, **143**:1399−1408 (2021). https://dx.doi.org/10.1021/jacs.0c10661.

[44] Yunan Wang, Feng Cao, Weiwei Lin, Fengyu Zhao, Jun Zhou, Song Li and Gaowu Qin. In situ synthesis of Ni/NiO composites with defect rich ultrathin nanosheets for highly efficient biomass-derivative selective hydrogenation. *J. Mater. Chem. A*, **7**:17834-17841 (2019). https://doi.org/10.1039/C9TA04487F.

[45] Beata Zielinska, Beata Michalkiewicz, Xuecheng Chen, Ewa Mijowska, and Ryszard J. Kalenczuk. Pd supported ordered mesoporous hollow carbon spheres (OMHCS) for hydrogen storage. *Chemical Physics Letters* **647**:14–19 (2016). http://dx.doi.org/10.1016/j.cplett.2016.01.036.

[46] S. Schaefer, V. Fierro, A. Szczurek, M.T. Izquierdo and A. Celzard. Physisorption, chemisorption and spill-over contributions to hydrogen storage. *International Journal of Hydrogen Energy* **41**: 17442-17452 (2016). https://doi.org/10.1016/j.ijhydene.2016.07.262.

[47] Kejun Zeng, T. Klassen, W. Oelerich, and R. Bormann. Thermodynamics of the Ni–H system. *Journal of Alloys and Compounds* **283**: 151-161 (1999). https://doi.org/10.1016/S0925-8388(98)00857-3.

[48] Andrea Centrone, Luigi Brambilla, and Giuseppe Zerbi. Adsorption of $H_2$ on carbon-based materials: A Raman spectroscopy study. *Physical Review B* **71**: 245406 (2005). DOI:10.1103/PhysRevB.71.245406.